\documentclass[
 reprint,
 amsmath,
 amssymb,
 aps,
 prl,
 superscriptaddress,
 floatfix,
 longbibliography
]{revtex4-2}

\usepackage{graphicx}
\usepackage{dcolumn}
\usepackage{bm}

\begin{document}


\title{Unified Theory of Quartz Tuning Fork Resonators}

\author{Hankyul Koh}
\affiliation{
Department of Physics and Astronomy,
College of Natural Sciences,
Seoul National University,
Seoul 08826, Korea
}

\author{Wonho Jhe}
\email{whjhe@snu.ac.kr}
\affiliation{
Department of Physics and Astronomy,
College of Natural Sciences,
Seoul National University,
Seoul 08826, Korea
}
\affiliation{
Multiscale Instruments Inc., Seoul 08510, Korea 
}





\date{\today}

\begin{abstract}
Quartz tuning forks, functioning as electrically driven piezoelectric resonators, have long served as exceptionally stable and widely adopted timing references in diverse domains of research and industry.
Yet, experimentally measured electrical resonance spectra often exhibit resonance evolutions that remain unexplained within existing theoretical descriptions.
Here we develop a unified continuum electromechanical modal framework that integrates piezoelectric electrodynamics, variational structural dynamics, and symmetry-selected electromechanical observability.
Our theory shows quantitative agreement with experimental results, demonstrating that electrical observability emerges not from the underlying mechanical eigenmodes alone.
The resulting framework unifies conventional coupled-oscillator, equivalent-circuit, and continuum descriptions within a single first-principles theory and provides a rigorous basis for precision electromechanical characterization.
\end{abstract}

\maketitle

Modern electronic and measurement systems increasingly rely on precise timing, synchronization, and resonant signal transduction \cite{Mengali1997Synchronization,Demir2000PhaseNoise}.
As physical systems of interest approach nanometer length scales \cite{Ekinci2005NEMS} and ultrafast timescales \cite{Emma2010LCLS}, even weak frequency shifts, modal distortions, or synchronization errors can generate substantial relative error in precision communication, synchronization, and metrology.
Among the most widely used self-sensing resonant elements are quartz-based piezoelectric resonators \cite{Cady1922PiezoElectricResonator}, including quartz tuning forks (QTFs) \cite{Karrai1995_PiezoTipSample,Giessibl1998_HighSpeedForceSensor,Grober2000ForceDetectionQTF}, owing to their high frequency stability, high quality factors, and direct electrical excitation and readout \cite{Mason1950Piezoelectric}.

Continuum theories of piezoelectricity establish the governing electroelastic dynamics of resonators through constitutive relations, quasi-electrostatic field equations, and elastic wave equations \cite{Mindlin1952PiezoPlates,Tiersten1963}, with the underlying eigenmodes and eigenfrequencies determined by the geometry, material anisotropy, and boundary conditions of the structure \cite{LandauLifshitz1986Elasticity}.
Electrode geometry can strongly influence electromechanical excitation and detection efficiency \cite{Coquin1967_SurfaceElectrodes}, while mode localization \cite{Spletzer2006ModeLocalizationMassSensing} and modal hybridization \cite{Haus1991CoupledModeTheory} can redistribute modal participation and electrical spectral weight among nearby resonances.
Despite this well-established continuum framework, the connection between the underlying electroelastic dynamics and experimentally observed electrical resonance spectra remains nontrivial.

In practice, QTFs are not perfectly symmetric and inevitably exhibit some degree of mass imbalance arising from fabrication tolerances, electrode nonuniformity, or probe attachment.
Under such perturbations, experimentally measured electrical spectra often exhibit complex evolution that differs markedly from that predicted by conventional coupled harmonic oscillator (CHO) \cite{Naber1999_dynamic,CastellanosGomez2009_QTF_Dynamics} descriptions.
This discrepancy suggests that the experimentally observed electrical response is a subtle manifestation of the underlying electroelastic dynamics rather than a direct representation of the mechanical eigenspectrum alone.
While Butterworth--Van Dyke (BVD) equivalent \cite{Butterworth1914,VanDyke1928} circuits provide efficient phenomenological descriptions of individual resonance peaks and finite-element simulations \cite{Allik1970PiezoFEM,Oria2013_FEA_QTF} can reproduce realistic geometries, material anisotropy, and mode shapes, neither framework by itself elucidates how continuum electromechanical dynamics are translated into experimentally observed electrical resonance spectra.

In this Letter, we formulate a unified continuum electromechanical modal theory for anisotropic piezoelectric solids from the electric enthalpy functional and derive a reduced electromechanical description by projecting continuum mechanical and electrode-defined electrostatic fields onto reduced modal subspaces.
The resulting first-principles theory directly connects continuum electromechanics to experimentally measured electrical resonance spectra through a unified electromechanical projection formalism.
Our theory provides a first-principles description of electromechanical observability, symmetry-breaking-induced mode splitting, and modal hybridization, while commonly used CHO models, BVD equivalent circuits, and qPlus-type single-prong configurations \cite{Giessibl1998_HighSpeedForceSensor} emerge naturally as reduced projections or limiting cases of the same continuum framework.

\begin{table*}[t]
\caption{\label{tab:related_models}
Comparison of existing QTF descriptions and their emergence as reduced limits of the present unified continuum electromechanical modal theory.}

\begin{ruledtabular}
\begin{tabular}{lllll}

\parbox[t]{0.15\textwidth}{Approach} &
\parbox[t]{0.16\textwidth}{Representative works} &
\parbox[t]{0.21\textwidth}{Strengths} &
\parbox[t]{0.22\textwidth}{Limitations} &
\parbox[t]{0.22\textwidth}{Emergence within the present theory}
\\[1pt]
\hline
\noalign{\vskip 2pt}

\parbox[t]{0.15\textwidth}{Single-oscillator models} &
\parbox[t]{0.16\textwidth}{Karrai--Grober \cite{Karrai1995_PiezoTipSample}} &
\parbox[t]{0.21\textwidth}{Simple resonance and piezoelectric response description} &
\parbox[t]{0.22\textwidth}{Electrode geometry and anisotropic coupling not explicit} &
\parbox[t]{0.22\textwidth}{Single-mode projection of the present framework}
\\[2pt]
\parbox[t]{0.15\textwidth}{Coupled-oscillator models} &
\parbox[t]{0.16\textwidth}{Naber \cite{Naber1999_dynamic}; Castellanos-Gomez et al.\ \cite{CastellanosGomez2009_QTF_Dynamics}} &
\parbox[t]{0.21\textwidth}{Captures symmetric and antisymmetric inter-prong coupling} &
\parbox[t]{0.22\textwidth}{Coupling and excitation symmetry imposed phenomenologically} &
\parbox[t]{0.22\textwidth}{Few-mode projection of the present theory}
\\[2pt]

\parbox[t]{0.15\textwidth}{qPlus models} &
\parbox[t]{0.16\textwidth}{Giessibl \cite{Giessibl1998_HighSpeedForceSensor}} &
\parbox[t]{0.21\textwidth}{Accurate description of fixed-prong force sensors} &
\parbox[t]{0.22\textwidth}{Special fixed-prong boundary condition} &
\parbox[t]{0.22\textwidth}{Constrained-motion limit of the general framework}
\\[2pt]

\parbox[t]{0.15\textwidth}{Finite-element analysis} &
\parbox[t]{0.16\textwidth}{Oria et al. \cite{Oria2013_FEA_QTF}} &
\parbox[t]{0.21\textwidth}{Realistic geometry and anisotropic mode shapes} &
\parbox[t]{0.22\textwidth}{Reduced dynamics not explicit} &
\parbox[t]{0.22\textwidth}{Numerical realization of continuum electroelasticity}
\\[2pt]

\parbox[t]{0.15\textwidth}{BVD/equivalent-circuit models} &
\parbox[t]{0.16\textwidth}{Lee et al. \cite{Lee2007_QTF_AFM}} &
\parbox[t]{0.21\textwidth}{Efficient impedance fitting and resonance extraction} &
\parbox[t]{0.22\textwidth}{Mode structure hidden in fitted parameters} &
\parbox[t]{0.22\textwidth}{Lumped limit of reduced dynamics}
\\[1pt]

\end{tabular}
\end{ruledtabular}
\end{table*}

Consider a linear anisotropic piezoelectric solid occupying a domain $\Omega$. Under infinitesimal strain \cite{LandauLifshitz1986Elasticity} and quasi-electrostatic conditions \cite{LandauLifshitz1984_continuous_media}, the coupled electroelastic dynamics are governed by the electric enthalpy density \cite{Tiersten1969,Yang1995VariationalPiezoelectric}
\begin{equation}
\mathcal H
=
\frac12
C^{E}_{ijkl}\varepsilon_{ij}\varepsilon_{kl}
-
e_{kij}\varepsilon_{ij}E_k
-
\frac12
\epsilon^{S}_{ij}E_iE_j ,
\label{eq:enthalpy}
\end{equation}
which yields the constitutive relations
\begin{equation}
\sigma_{ij}
=
C^{E}_{ijkl}\varepsilon_{kl}
-
e_{kij}E_k,
\qquad
D_i
=
e_{ijk}\varepsilon_{jk}
+
\epsilon^{S}_{ij}E_j,
\label{eq:constitutive}
\end{equation}
where $C^E_{ijkl}$ is the elastic stiffness tensor at constant electric field $E_i$, $\varepsilon_{ij}$ the infinitesimal strain tensor, $e_{kij}$ the piezoelectric tensor, $\epsilon^S_{ij}$ the dielectric tensor, $\sigma_{ij}$ the stress tensor, and $D_i$ the electric displacement field.

Under the quasi-electrostatic approximation, the electric field is determined instantaneously by the electrode boundary conditions through Gauss' law \footnote{Since the mechanical resonance frequency of QTFs lies in the kHz range while electromagnetic propagation occurs on much faster timescales, the electric field is assumed to adjust instantaneously to the electrode boundary conditions \cite{Tiersten1963}}.
To obtain a reduced electromechanical description suitable for resonance analysis, we project the continuum fields onto mechanical and electrode-defined electrostatic basis functions.
Repeated indices imply summation unless otherwise stated.
The mechanical displacement $u_i$ and electrostatic potential $\phi$ are expanded as
\begin{equation}
u_i(\mathbf x,t)
=
q_\alpha(t)\widehat u_i^{(\alpha)}(\mathbf x),
\qquad
\phi(\mathbf x,t)
=
V_\ell(t)\widehat\phi^{(\ell)}(\mathbf x),
\label{eq:projection}
\end{equation}
where $q_\alpha$ denotes the generalized coordinate associated with the mechanical basis function $\widehat u_i^{(\alpha)}$, and $\widehat\phi^{(\ell)}$ the electrostatic basis potential associated with unit voltage applied to the electrode $\ell$ with all other electrodes grounded. Notice that the electrode geometry determines not only the applied voltage distribution, but also the spatial symmetry and modal selectivity of the electromechanical excitation field.
For the QTF geometry considered below, $\widehat u_i^{(\alpha)}$ is constructed from Timoshenko-type flexural \cite{Timoshenko1921,Cowper1966} and torsional kinematics including Saint--Venant warping \cite{Lekhnitskii1963}.

Using Eq.~\eqref{eq:projection} together with $E_i=-\partial_i\phi$,
\begin{equation}
E_i(\mathbf x,t)
=
V_\ell(t)\widehat E_i^{(\ell)}(\mathbf x),
\qquad
\widehat E_i^{(\ell)}
=
-\partial_i\widehat\phi^{(\ell)} .
\label{eq:eprojection}
\end{equation}
The strain field is similarly expanded as
\begin{equation}
\varepsilon_{ij}(\mathbf x,t)
=
q_\alpha(t)
\widehat\varepsilon_{ij}^{(\alpha)}(\mathbf x),
\qquad
\widehat\varepsilon_{ij}^{(\alpha)}
=
\frac12
\left(
\partial_i\widehat u_j^{(\alpha)}
+
\partial_j\widehat u_i^{(\alpha)}
\right).
\label{eq:strainprojection}
\end{equation}
The continuum mechanical weak form for virtual displacements $\delta u_i$ takes the form
\begin{equation}
\int_\Omega
\sigma_{ij}\delta\varepsilon_{ij}
\,d\Omega
+
\int_\Omega
\rho \ddot u_i \delta u_i
\,d\Omega
=
\int_\Omega
f_i\delta u_i
\,d\Omega
+
\int_{\Gamma_t}
\bar t_i\delta u_i
\,dA ,
\label{eq:mechweak}
\end{equation}
with $\delta\varepsilon_{ij}=\frac12\left(\partial_i\delta u_j+\partial_j\delta u_i\right).$
Substituting Eqs.~\eqref{eq:projection}--\eqref{eq:strainprojection} into Eq.~\eqref{eq:mechweak} and projecting onto the mechanical basis $\delta u_i=\delta q_\alpha \widehat u_i^{(\alpha)}$ yields the reduced electromechanical equation of motion, given by
\begin{equation}
M_{\alpha\beta}\ddot q_\beta
+
D_{\alpha\beta}\dot q_\beta
+
K_{\alpha\beta}q_\beta
=
\Theta_{\alpha\ell}V_\ell ,
\label{eq:reduced}
\end{equation}
where $M_{\alpha\beta}$, $D_{\alpha\beta}$, and $K_{\alpha\beta}$ are the projected inertia, damping, and stiffness matrices of the mechanical subspace, respectively, while $\Theta_{\alpha\ell}$ is the reduced electromechanical coupling matrix.
Notice that Eq.~\eqref{eq:reduced} defines a reduced electromechanical projection structure from which conventional CHO models emerge naturally as low-dimensional projections of the underlying continuum electromechanical modal theory.

The electromechanical coupling matrix $\Theta_{\alpha\ell}$ arises from the piezoelectric interaction term in Eq.~\eqref{eq:enthalpy}.
Using Eqs.~\eqref{eq:eprojection} and \eqref{eq:strainprojection},
\begin{equation}
-e_{kij}\varepsilon_{ij}E_k
=
-q_\alpha V_\ell
e_{kij}
\widehat\varepsilon_{ij}^{(\alpha)}
\widehat E_k^{(\ell)} .
\label{eq:interaction}
\end{equation}
Therefore, the coefficient of the $q_\alpha V_\ell$ term defines
\begin{equation}
\Theta_{\alpha\ell}
=
\int_\Omega
e_{kij}
\widehat\varepsilon_{ij}^{(\alpha)}
\widehat E_k^{(\ell)}
\,d\Omega .
\label{eq:theta}
\end{equation}
Notice that Eq.~\eqref{eq:theta} shows that the operator $\Theta_{\alpha\ell}$ is determined by the electromechanical overlap between the mechanical strain field and the electrode-defined excitation field and therefore governs the modal selectivity of electrical excitation.

Let us now consider the electrical readout generated by the mechanically induced polarization charge on the electrode surface $\Gamma_\ell$.
Calculating the flux of $D_i$ in Eq.~\eqref{eq:constitutive} over $\Gamma_\ell$, together with Eqs.~\eqref{eq:eprojection} and \eqref{eq:strainprojection}, we obtain
\begin{equation}
Q_\ell
=
q_\alpha
\int_{\Gamma_\ell}
n_i e_{ijk}
\widehat\varepsilon_{jk}^{(\alpha)}
\, dA
+
V_r
\int_{\Gamma_\ell}
n_i \epsilon^S_{ij}
\widehat E_j^{(r)}
\, dA ,
\label{eq:chargeexpanded}
\end{equation}
where $n_i$ denotes the outward surface normal.
Here, the first term represents the mechanically induced piezoelectric surface charge, while the second term corresponds to the purely dielectric electrode response.
In the operator form, Eq.~\eqref{eq:chargeexpanded} is expressed as
\begin{equation}
Q_\ell
=
H_{\ell\alpha} q_\alpha
+
C_{\ell r} V_r ,
\label{eq:charge}
\end{equation}
where
\begin{equation}
H_{\ell\alpha}
=
\int_{\Gamma_\ell}
n_i e_{ijk}
\widehat\varepsilon_{jk}^{(\alpha)}
\, dA
\label{eq:Hdef}
\end{equation}
defines the reduced piezoelectric readout operator.
Equation~\eqref{eq:charge} shows that electrical readout is governed by the projection operator $H_{\ell\alpha}$ in direct analogy with electromechanical excitation through $\Theta_{\alpha\ell}$.
Note that the matrix element $H_{\ell\alpha}$ quantifies how strongly the mechanical basis component $\alpha$ generates the measurable charge on the electrode $\ell$. 
The second contribution in Eq.~\eqref{eq:charge} defines the reduced dielectric capacitance operator whose matrix element is obtained as
\begin{equation}
C_{\ell r}
=
\int_\Omega
\epsilon^S_{ij}
\widehat E_i^{(\ell)}
\widehat E_j^{(r)}
\, d\Omega ,
\label{eq:capacitance}
\end{equation}
which follows from projection of the dielectric contribution in Eq.~\eqref{eq:constitutive}.
Notice that the capacitance matrix $C_{\ell r}$ represents the continuum generalization of the static dielectric capacitance $C_0$ that appears in the BVD-type equilvalent-circuit models of piezoelectric resonators. Interestingly, the capacitance emerges directly from the continuum electromechanical modal formulation, without the need for a phenomenological circuit-level introduction.

Concerning the operators $\Theta_{\alpha\ell}$ and $H_{\ell\alpha}$, they are not independent, but arise as adjoint projections of the same electroelastic bilinear form in Eq.~\eqref{eq:interaction}.
Consequently, the two operators are related through $H=\Theta^\dagger$, which reflects electromechanical reciprocity \cite{Coquin1967_SurfaceElectrodes,Auld1973AcousticFields}. Electrical excitation and electrical observability are therefore governed by the same symmetry-selected electroelastic overlaps. For conventional QTF electrodes, the electrode-defined electric field is odd under prong exchange. Consequently, parity symmetry permits finite electroelastic overlap only for antisymmetric strain fields, yielding the leading-order selection rule that antisymmetric modes are electrically observable whereas symmetric modes remain electrically suppressed.

Let us now discuss the relation of our operator formalism to the measured electrical signal.
Projection onto the mechanical eigenmodes yields the effective electromechanical excitation and readout amplitudes
\begin{equation}
\Theta^{\mathrm{eff}}_{n\ell}
=
v_\alpha^{(n)}
\Theta_{\alpha\ell},
\qquad
H^{\mathrm{eff}}_{\ell n}
=
H_{\ell\alpha}
v_\alpha^{(n)},
\end{equation}
where $v_\alpha^{(n)}$ denotes the mass-normalized eigenvector of mode $n$.
The quantity $\Theta^{\mathrm{eff}}_{n\ell}$ measures how strongly the electrode $\ell$ excites the mode $n$, while $H^{\mathrm{eff}}_{\ell n}$ determines how efficiently that mode generates measurable electrical signal on the readout electrode.
Combining electromechanical excitation, modal dynamics, and electrical readout yields the measured electrical response
\begin{equation}
\tilde I_{\mathrm{meas}}(\omega)
=
-i\omega
\sum_n
\frac{
s_\ell
H^{\mathrm{eff}}_{\ell n}
\Theta^{\mathrm{eff}}_{nr}
}{
\omega_n^2-\omega^2-i\omega\gamma_n
}
\tilde V_r ,
\label{eq:response}
\end{equation}
where $\omega_n$ and $\gamma_n$ denote the modal resonance frequency and damping rate, respectively, and $s_\ell$ specifies the measurement configuration.

Equation~\eqref{eq:response}, our main result, shows that a mechanically resonant mode contributes to the measured electrical spectrum only when both the excitation and readout overlaps are nonvanishing.
Notice that Eq.~\eqref{eq:response} separates the measured electrical spectrum into two distinct structures: the electromechanical numerator governs modal observability, whereas the pole denominator determines the mechanical resonance frequencies and linewidths.
For convenience, we define
\begin{equation}
\mathcal V_n
\equiv
\left|
H^{\mathrm{eff}}_{\ell n}
\Theta^{\mathrm{eff}}_{n r}
\right|,
\label{eq:visibility}
\end{equation}
which measures the combined excitation and readout overlap governing the electrical visibility of mode $n$ in the measured signal.
The visibility $\mathcal V_n$ determines whether a mechanical pole contributes appreciably to the measured electrical spectrum, but does not by itself determine how the underlying mechanical poles evolve under symmetry breaking, mass loading, or other perturbations.

\begin{figure*}
\includegraphics[width=1.0\textwidth]{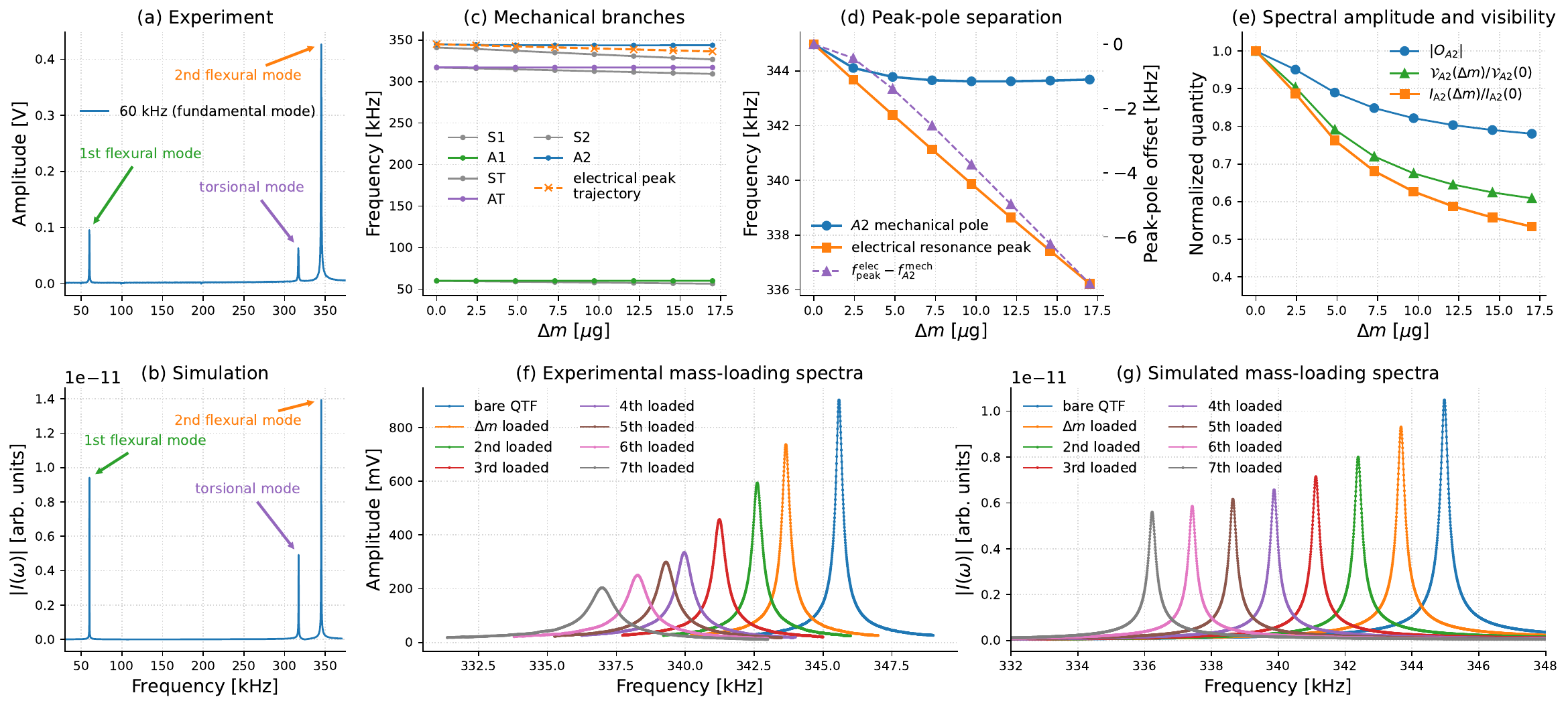}
\caption{
Experimental and theoretical evolution of electromechanically observable resonance branches in QTF under asymmetric mass loading.
(a), (b) Experimental and simulated wide-band electrical spectra for representative QTF devices.
(c) Mechanical pole evolution under progressive mass perturbation with the trajectory of the electrically observed resonance peak.
(d) Comparison between the tracked A2-like (second antisymmetric flexural) mechanical pole and the observed electrical resonance peak.
(e) Evolution of the modal overlap amplitude, electromechanical visibility, and normalized electrical peak amplitude.
(f), (g) Experimental and simulated spectra near the second antisymmetric flexural branch under progressive mass loading.
In the simulations, hydrodynamic loading was incorporated using Sader's rectangular-cantilever formulation \cite{Sader1998FrequencyResponse,Sader1999Calibration}.
}
\label{fig:sweep2}
\end{figure*}

For realistic QTF devices, perturbations due to mass imbalance
\cite{Cleveland1993AFM,CastellanosGomez2009_QTF_Dynamics,CastellanosGomez2011_Dissipation,Kim2014_QPlus_Stiffness}
and electrode nonuniformity
\cite{Coquin1967_SurfaceElectrodes,Tiersten1969}
affect the mechanical dynamics as well as the electromechanical coupling structure in Eq.~\eqref{eq:reduced},
\begin{equation}
M=M^{(0)}+\delta M,\;
\qquad
K=K^{(0)}+\delta K.
\end{equation}
Such imperfections generally produce symmetry-broken pole splitting and modal hybridization across the coupled branches, consistent with the branch splitting predicted by conventional CHO descriptions. Experimentally, however, the observed resonance response is often dominated by a single antisymmetric-like resonance branch despite the presence of multiple nearby mechanical poles [Fig.~\ref{fig:sweep2}(c)].
Under perturbation, the reduced eigenmodes satisfy
\begin{equation}
K_{\alpha\beta}v_\beta^{(n)}
=
\omega_n^2
M_{\alpha\beta}v_\beta^{(n)} .
\end{equation}
To leading order, the pole displacement is governed by
\begin{equation}
\delta\omega_n^2
=
(v^{(n)})^T
(
\delta K
-
\omega_n^2\delta M
)
v^{(n)},
\label{eq:poleshift}
\end{equation}
while the corresponding eigenvector correction becomes
\begin{equation}
\delta v_\alpha^{(n)}
=
\sum_{m\neq n}
v_\alpha^{(m)}
\frac{
(v^{(m)})^T
(
\delta K
-
\omega_n^2\delta M
)
v^{(n)}
}{
\omega_n^2-\omega_m^2
}.
\label{eq:hybridization}
\end{equation}
More generally, in the presence of damping, the perturbation acts on the complex pole structure of the reduced electromechanical system, leading to resonance-frequency shifts, linewidth changes, and mode hybridization, while Eqs.~\eqref{eq:poleshift} and \eqref{eq:hybridization} provide the corresponding leading-order description in the weak-damping limit.

For localized mass imbalance, for example, the dominant contribution comes from $\delta M_{\alpha\beta}$. As a result, the diagonal mass perturbation element $(v^{(n)})^T\delta M\,v^{(n)}$ governs the mechanical pole displacement, whereas the off-diagonal overlaps control modal hybridization between nearby branches.
Since the reduced modes satisfy a generalized eigenvalue problem, their natural orthogonality is defined by the mass metric, $(v_m^{(0)})^T M^{(0)}v_n^{(0)}=\delta_{mn}$.
We therefore quantify the modal overlap amplitude under perturbation by
\begin{equation}
O_n
=
\left|
\left(v_n^{\rm load}\right)^T
M^{(0)}
v_n^{(0)}
\right|.
\label{eq:overlap}
\end{equation}
Since the measured current depends on the product of the electromechanical excitation and readout amplitudes, the modal overlap amplitude contributes to both projections.
To leading order,
\begin{equation}
(v_n^{\rm load})^T\Theta
\simeq
O_n (v_n^{(0)})^T\Theta,
\qquad
H^T v_n^{\rm load}
\simeq
O_n H^T v_n^{(0)} .
\end{equation}
Consequently, the modal contribution to the measured current scales approximately as
\begin{equation}
I_n(\omega)
\simeq
O_n^2
I_n^{(0)}(\omega).
\label{eq:spectral_weight_scaling}
\end{equation}
The quantity $O_n^2$, referred to here as the modal spectral weight, provides a leading-order estimate of the reduction in observable electromechanical response arising solely from eigenvector hybridization.

Figure~\ref{fig:sweep2} compares the experimentally measured and theoretically simulated electrical spectra under progressive asymmetric mass loading.
The spectral evolution is interpreted through the interplay of pole displacement [Eq.~\eqref{eq:poleshift}], modal hybridization [Eq.~\eqref{eq:hybridization}], modal overlap amplitude [Eq.~\eqref{eq:overlap}], electromechanical visibility [Eq.~\eqref{eq:visibility}], and modal spectral-weight scaling [Eq.~\eqref{eq:spectral_weight_scaling}].
As shown in Figs.~\ref{fig:sweep2}(a) and \ref{fig:sweep2}(b), the theory quantitatively reproduces the experimentally observed wide-band electrical spectrum of the mass-loaded 60~kHz quartz tuning fork, including the evolution of the dominant electrically observable resonance branch under increasing asymmetric loading.
Figures~\ref{fig:sweep2}(c) and \ref{fig:sweep2}(d) show that the observed electrical resonance trajectory does not simply follow the underlying mechanical pole evolution.
The experimentally observed resonance frequency decreases from about 346~kHz to 337~kHz, while the measured quality factor decreases from about 1760 to 428.
For a damped harmonic oscillator, $f_d=f_n\sqrt{1-\zeta^2}$ with $\zeta=1/(2Q)$, yielding a damping-induced frequency correction below 0.3~Hz.
The observed frequency change of approximately 8.6~kHz therefore predominantly reflects loading-induced changes in the mechanical pole structure and modal composition rather than damping effects alone.
Figure~\ref{fig:sweep2}(e) shows that the measured peak suppression cannot be inferred from pole evolution alone.
The calculated loss of electromechanical observability accounts for about $62.5\%$ of the observed $80\%$ peak suppression, corresponding to about $50$ percentage points of the original amplitude, whereas the included Sader hydrodynamic contribution produces only about a $0.03\%$ reduction and therefore cannot explain the residual loss.
The remaining suppression suggests additional loading-dependent dissipation, such as anchor loss at the base and center-of-mass motion induced by symmetry breaking \cite{Shaskey2024StronglyCoupledQTF}, whose quantitative treatment requires a dedicated damping model and will be the subject of future work.
Finally, Figs.~\ref{fig:sweep2}(f) and \ref{fig:sweep2}(g) compare the detailed resonance evolution observed experimentally and predicted by the theory.
The agreement demonstrates that the theory captures the dominant spectral trends, including the progressive frequency shift and amplitude reduction of the electrically visible resonance branch, while maintaining a direct connection between experimentally measurable electrical quantities and the underlying continuum electromechanical eigenstructure.

In conclusion, we have developed a continuum electromechanical modal theory that establishes a direct connection between anisotropic piezoelectric field equations, electrode geometry, and experimentally measured electrical resonance spectra.
By revealing how continuum eigenmodes become electrically observable through electrode-defined excitation and readout, the theory provides a unified description of resonance visibility, mode hybridization, and symmetry-breaking-induced spectral evolution that cannot be understood from mechanical pole dynamics alone.
Beyond quartz tuning forks, our theory is expected to be broadly applicable to multimode piezoelectric resonators used in electromechanical sensing, precision timing, synchronization, and electrically calibrated force measurements.

\begin{acknowledgments}
The author is particularly indebted to Dr. Joon-Hyuk Ko for his valuable suggestions and help on this work. This work was supported by grants from the National Research Foundation of Korea (No. 2016R1A3B1908660) and under the DeepTech TIPS Program (No. RS-2025-25461650). 
\end{acknowledgments}

\paragraph*{Data availability.}
The experimental data and simulation code supporting the findings of this study will be made available in a public repository upon publication. Until then, they are available from the corresponding author upon reasonable request.

\bibliography{references}
\bibliographystyle{apsrev4-2} 


\end{document}